\documentclass{article}
\usepackage[utf8]{inputenc}
\UseRawInputEncoding
\usepackage{geometry}
\usepackage{generic}
\usepackage{cite}
\usepackage{amsmath,amssymb,amsfonts}
\usepackage{algorithmic}
\usepackage{graphicx}
\usepackage{textcomp}
\usepackage{colortbl}
\usepackage{lastpage, hyperref}
\usepackage{textgreek}
\usepackage{subcaption}
\usepackage{adjustbox}
\usepackage{caption}
\usepackage{nccmath}
\usepackage{authblk}
\DeclareCaptionLabelFormat{AppendixTables}{A.#2}
\newcommand{\argmax}{\arg\!\max}

\title{Towards Fast Single-Trial Online ERP based Brain-Computer Interface using dry EEG electrodes and neural networks: a pilot study}

\author{Okba Bekhelifi*, Nasr-Eddine Berrached}

\affil{Intelligent Systems Research Laboratory, Electronics department, Université des Sciences et de la Technologie d’Oran Mohamed Boudiaf (USTO-MB), El Mnaouar, BP 1505, Bir El Djir 31000, Oran, Algeria.}

\affil[]{\textit {\{okba.bekhelifi, nasreddine.berrached\}@univ-usto.dz}}

\begin{document}

\date{}
\maketitle

\section*{Abstract}
Speeding up the spelling in event-related potentials (ERP) based Brain-Computer Interfaces (BCI) requires eliciting strong brain responses in a short span of time, as much as the accurate classification of such evoked potentials remains challenging and imposes hard constraints for signal processing and machine learning techniques. Recent advances in stimulus presentation and deep learning showcased a promising direction in significantly improving the efficacy of those systems, in this study we propose the combination of colored inverted face stimulation with classification using convolutional neural networks in the hard settings of dry electrodes and fast flashing single-trial ERP-based BCI.
The high online accuracy achieved, with two subjects passing the 90\% correct symbol detection bar and a transfer rate above 60 bits per minute, demonstrates the approach’s potential in improving the practicality of ERP based BCIs.              
\section{Introduction}
A Brain-Computer interface (BCI) provides an alternative to usual muscular control for communication and interaction with the environment, although for such systems the main user target is people suffering neurodegenerative diseases
\cite{1}, healthy users operate them for new control experiences \cite{2}. Their applications vary over a wide range including, but not limited to, word spelling \cite{3}, powered wheelchair control \cite{4}, drone control \cite{5}, neuroprostheses and neurorehabilitation \cite{6}. An electroencephalography (EEG) based BCI relies on the translation of the brain activity measured by electrodes placed on the surface of the scalp \cite{7}. The safety, relative ease of use, high temporal resolution and portability are the key characteristics that made this technique the most used brain recording modality for BCI. The possible signal captured by EEG determines the components that generate distinct features used to differentiate the control options made available by the system \cite{8}. Additionally, depending on the triggering mechanism of the control signal, whether the user voluntary elicits the brain response or it is evoked by an external stimulus, there are two families of neural responses: endogenous and exogenous. In the former category the most common task employed is motor imagery (MI), in which the imagined movements of body parts such as hands and feet modulates the EEG rhythms acquired from sensorimotor cortex \cite{9}. Conversely, in the external stimulation category, visual stimulation is widely adopted, and can be achieved by means of presenting an infrequent stimulus to the subject in a fixed duration, this will elicit specific positive and negative peaks in the signal amplitude, occurring at known intervals after the stimulus onset, these deflections are called event-related potentials (ERPs) and are associated with sensory, cognitive, affective, and motor processes \cite{10}. Another popular visual stimulation technique is the steady-state visual evoked potentials (SSVEPs), where constant frequency flickering is presented to the user, which elicits in the EEG signal quasi-sinusoidal oscillations consistent with the flickering rate. The resulting signal will have higher frequency power over the occipital and parietal-occipital areas \cite{11}.  This study focus on ERP-based BCIs.

The basis of most ERP-based BCIs was introduced by \cite{3}, in their seminal work, a 6x6 letters matrix was displayed to subjects as stimulus, each row and column is flashed in a random order and subjects are asked to attend to the desired character and silently count the number of flashes. The row and column of the attended item should elicit the ERP component, and thus the desired character can be determined by intersection of the row and column with the highest amplitude. Here, the sought ERP component is the P300, a positive deflection in EEG around 300 ms after stimulus onset \cite{12}. The P300 is related to the interaction with a deviant stimuli and is largest over central and parietal areas \cite{13}, other prominent ERP components have been observed to be modulated in the P300 paradigm, In particular early responses N100, P100, P200 and N200.  In the same manner of P300, the N200 is associated with the processing of deviant stimuli, whereas the visual P100, N100 and P200 are associated with attention and spatial location \cite{14}. The major drawback of ERP-based BCI is the reliance on averaging several flashes to achieve clear amplitude peaks and acceptable performance, which increases the time required to output a single command making the system slow and unpractical.

In an attempt to remedy this issue, a successful approach has been introduced and had launched a series of investigations in the design of optimal stimulus. Instead of evoking the ERPs by means of flashing characters, \cite{15} proposed to use familiar faces, since the perception of faces involves several ERPs alongside the P300,  the specific face N170 and N400f \cite{16}. The resulting ERPs showed more robustness and yielded higher performance. \cite{17} tested the face-sensitive potentials against different configurations and reported that inverted faces produce better N170, vertex positive potential (VPP) and P300 components than upright face. \cite{18} Compared six stimulation conditions involving stimulus movement, neutral face stimulus and facial expression stimulus, their findings corroborated the superiority of face based ERPs over the traditional ones in all configurations. 

Another line of inquiry investigated the modification of the stimulus chromatic property, \cite{19} tested a green familiar face against a natural face, the addition of color was beneficial for ERP components and spelling accuracy. Based on this study results the work in \cite{20} compared the green face with two other colors blue and red, they reported superior performance achieved with the red face stimulus. \cite{21} went further with red face stimulus and experimented with red face and colored block shapes, the red face with white rectangle showed the highest online accuracy. These studies did not take into account the repetition effect of using the same face, in \cite{22} the authors proposed a multi-faces approach which evoked more stable components and yielded better performance compared to the established single face approach.

Traditionally, the electrodes used to capture EEG are made of silver/silver chloride, regardless of the decent signal quality they provide, a conductive connection with the skin is required, In this case additive gel is applied between the electrode and the scalp. Mounting the electrodes takes long time, subsequently, checking the gel regularly and hair washing after experiments are mandatory. These manipulations undermines the usability and comfort of such electrodes. Recently, alternative dry electrodes were proposed for quicker setup time, longer and convenient use. Unfortunately, they proved less reliable than their wet counterparts in terms of signal quality and performance \cite{23}. A substantial drop in performance was observed in an ERP-based study \cite{24} between wet and dry electrodes, suggesting for a more careful handling of the latter type and for applying more advanced signal processing methods for enhancing the signal-to-noise ratio.
 
The long sought pinnacle in ERP research is to attain correct target detection with a single trial. Parallel to optimizing the stimulus presentation and timing, accurate and robust signal processing and classification algorithms were developed \cite{25}, yet no satisfying approach has been able to withstand the high noise and non-stationarity embedded in the EEG signal. Deep neural networks have demonstrated superiority in many fields over other machine learning algorithms \cite{26}, their adoption in the BCI field was delayed due to the limited size of publicly available datasets, the difficulty of collecting larger ones and the failure to train models in similar settings. The seminal works of \cite{27} and \cite{28} proved the  superiority of small-to-medium size neural networks against state of the art methods in multiple BCI paradigms,  mostly in offline settings. For a review see \cite{29}.  
 
In online settings, the study in \cite{30} was the first to use a Convolutional neural network (CNN) to classify ERPs online, they used a small 3-layers CNN to control a dual M-VEP stimuli BCI, which proved significantly higher than standard approaches. \cite{31} explored the use of a 6-layers CNN in a rapid serial visual presentation (RSVP) experiment to study BCI illiteracy \cite{32} in that paradigm. In \cite{33} the authors combined Fast-Fourier Transform (FFT) with a 1D CNN to achieve a lightweight single channel SSVEP based BCI. In \cite{34} a virtual reality game controlled by MI for neuro-rehabilitation employed a larger CNN having 4 convolutional blocks followed by 6 fully connected layers, allowed a real-time control in a duration of 0.5 second. 
 
The fastest asynchronous non-invasive BCI reported to date \cite{35} used a CNN for classification in a code-modulated VEP (C-VEP) paradigm and achieved astonishing results, however the technical complexity of the solution limits its chances of replication, it required using 3 computers, among them one equipped with 4 GPUs for model training and inference. To handle the inherent small data size issue the work of \cite{36} relied on a 5-Layer Bayesian CNN (BCNN) for an ERP based Gomoku game BCI, the BCNN outperformed both its conventional CNN counterparts and classical methods.
  
In this study, for the goal of producing a fast reliable BCI, we investigate the combination of the following elements. First, the fusion of face configural processing (face inversion) introduced by \cite{17} with color configuration \cite{20} to construct an improved inverted red face ERP stimulus. Second, a rapid system throughput by means of fast flashing and the hard constraint of a single sequence for producing a decision. Third, compact CNNs to surpass the dry electrodes low signal-to-noise ratio (SNR) and enable real-time processing in limited computational environment. We demonstrate the feasibility of such a system tested in online spelling experiments and report both the limitations and possible ways of improvement for a more stable control. 

\begin{table}[htp!]
\centering
\begin{tabular}{|lllllll|} 
\hline
Questionnaire 1 & \multicolumn{6}{l|}{~}                                                                                                                                       \\ 
\hline
\multicolumn{7}{|l|}{Personal Information}                                                                                                                                     \\ 
\hline
1               & \multicolumn{6}{l|}{Age}                                                                                                                                     \\
2               & \multicolumn{6}{l|}{Gender (Male = 0, Female = 1)}                                                                                                           \\
3               & \multicolumn{6}{l|}{BCI experience (number of experiences; naive = 0)}                                                                                       \\
4               & \multicolumn{6}{l|}{Right-handed = 0, Left-handed = 1, Ambidexter = 2}                                                                                       \\ 
\hline
\multicolumn{7}{|l|}{Physiological
  and psychological~}                                                                                                                       \\ 
\hline
1               & \multicolumn{6}{l|}{\begin{tabular}[c]{@{}l@{}}How long have you slept? \\(1-4 h = 1, 5-6 h = 2, 6-7 h = 3, 7-8 h = 4, \textgreater{}8 h = 5)\end{tabular}}  \\ 
\hline
2               & \multicolumn{6}{l|}{\begin{tabular}[c]{@{}l@{}}Did you drink coffee in the last 24 hours? \\(in hours since last consumption; none = 0)\end{tabular}}        \\ 
\hline
3               & \multicolumn{6}{l|}{\begin{tabular}[c]{@{}l@{}}Did you drink alcohol in the last 24 hours? \\(in hours since last consumption; none = 0)\end{tabular}}       \\ 
\hline
4               & \multicolumn{6}{l|}{\begin{tabular}[c]{@{}l@{}}Did you smoke in the last 24 hours? \\(in hours since last consumption; none = 0)\end{tabular}}               \\ 
\hline
5               & Condition check list & Low & ~ & ~ & ~ & High                                                                                                                \\ 
\hline
~               & Comfort              & 1   & 2 & 3 & 4 & 5                                                                                                                   \\
~               & Motivation           & 1   & 2 & 3 & 4 & 5                                                                                                                   \\
~               & Eye fatigue          & 1   & 2 & 3 & 4 & 5                                                                                                                   \\
~               & Drowsiness           & 1   & 2 & 3 & 4 & 5                                                                                                                   \\
~               & Physical condition   & 1   & 2 & 3 & 4 & 5                                                                                                                   \\
~               & Mental condition     & 1   & 2 & 3 & 4 & 5                                                                                                                   \\
\hline
\end{tabular}
\caption{Before experiment questionnaire}
\end{table}

\begin{table}[htp!]
\centering
\arrayrulecolor{black}
\begin{tabular}{!{\color{black}\vrule}l|lllllp{2cm}!{\color{black}\vrule}} 
\hline
\multicolumn{7}{!{\color{black}\vrule}l|}{Questionnaire 2}                                                                                                                             \\ 
\hline
Phase & \multicolumn{6}{l!{\color{black}\vrule}}{Calibration or online}                                                                                                                \\ 
\hline
1     & Condition check list                                                                                  & Low & ~ & ~ & ~ & High                                                 \\ 
\cline{1-1}\arrayrulecolor{black}\cline{2-6}\arrayrulecolor{black}\cline{7-7}
~     & Comfort                                                                                               & 1   & 2 & 3 & 4 & 5                                                    \\
~     & Motivation                                                                                            & 1   & 2 & 3 & 4 & 5                                                    \\
~     & Concentration                                                                                         & 1   & 2 & 3 & 4 & 5                                                    \\
~     & Eye fatigue                                                                                           & 1   & 2 & 3 & 4 & 5                                                    \\
~     & Drowsiness                                                                                            & 1   & 2 & 3 & 4 & 5                                                    \\
~     & Physical condition                                                                                    & 1   & 2 & 3 & 4 & 5                                                    \\
~     & Mental condition                                                                                      & 1   & 2 & 3 & 4 & 5                                                    \\ 
\cline{1-1}\arrayrulecolor{black}\cline{2-6}\arrayrulecolor{black}\cline{7-7}
2     & Difficulty                                                                                            & 1   & 2 & 3 & 4 & 5                                                    \\ 
\hline
3     & \multicolumn{1}{p{2.5cm}!{\color{black}\vrule}}{Did you ever doze off or fall asleep during the experiment?} & \multicolumn{5}{l!{\color{black}\vrule}}{(number of times; none = 0)}  \\ 
\hline
4     & \multicolumn{1}{l!{\color{black}\vrule}}{Missed attempts}                                             & \multicolumn{5}{l!{\color{black}\vrule}}{(number; none = 0)}           \\ 
\hline
5     & \multicolumn{1}{p{2.5cm}!{\color{black}\vrule}}{Expected accuracy for this experiment (\%)}                  & \multicolumn{5}{l!{\color{black}\vrule}}{~}                            \\
\hline
\end{tabular}
\caption{After experiment questionnaire}
\end{table}

\section{Materials and methods}
\subsection{Participants}
6 healthy males participated in the experiments (age: mean 27.83, STD 3.33, range 23-32. 5 right-handed, 1 left-handed), all participants had normal or corrected-to-normal vision. Except for S1 all subjects had no experience in BCI control, all the experiments were conducted in accordance with standards of the Declaration of Helsinki (World Medical Association). Prior to each session, subjects filled information questionnaire form, received explanation on the study and signed a written informed consent. We followed the same forms reported in \cite{37} to record subjects’ personal information and both physiological and psychological states. Table 1 and Table 2 list the items and checklist to be filled before and after the experiments respectively.

\subsection{Experimental stimuli and paradigm}
Participants were seated in comfortable chair 70 cm approximately away from a laptop 15.6 inch LCD monitor (60 Hz refresh rate, 1366x768 screen resolution). Eight directions plus a stop command were presented in a 3x3 matrix to simulate the control of a powered wheelchair or a mobile robot. A red colored inverted face of the Algerian author Malek Bennabi wearing glasses were used as stimulus, with 124 x 157 pixels size. Figure 1 illustrates the stimulation layout. Icons were flashed with the face in a Single-Character (SC) fashion in a random order. 
\linebreak
Each subject completed a single run within the same day, a run was divided in two separate sessions, the first session is dedicated for calibration in which data were acquired for training the classifier and no feedback was presented,  whereas the second session consisted of two consecutive Copy-Spelling online testing phases where classifier’s decision was presented as feedback.

\begin{figure}[t]
  \includegraphics[width=\linewidth]{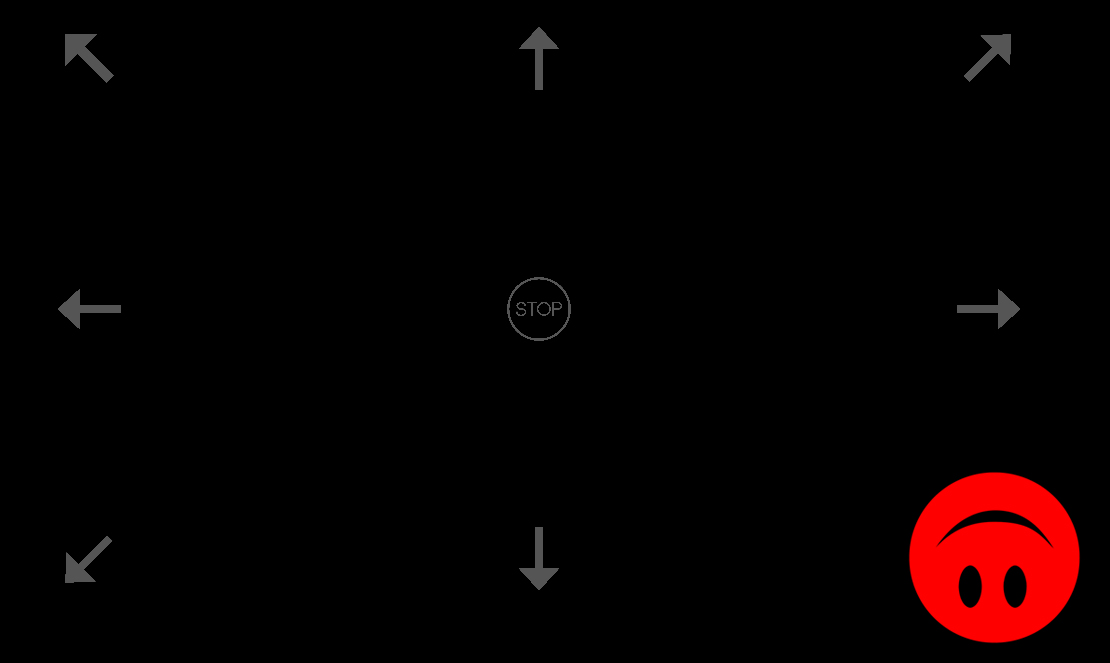}
  \caption{Stimulation layout with 9 commands and a single flash with red inverted face covering an icon during the stimulation period. For copyrights infringement issues we replaced the face stimulus used in the experiment by the inverted red dummy face in this figure.}
  \label{fig:1}
\end{figure}

The online sessions were separated by few minutes to give the subjects rest and test the classifier’s performance against inter-session variability.

In the calibration phase, at the start of each trial, a cue in the form of a yellow square first highlighted the target icon for 500ms, after that the flashing sequence starts, with a stimulus presentation duration of 40ms and inter-stimulus interval (ISI) of 70ms for a single flash (110ms stimulus onset asynchrony, SOA). A trial consists of nine consecutive flashes, and a sequence is the repetition of the trial 10 times. To avoid double target flashes in the random stimulation sequence, the distance between two similar flashes was set to a minimum of two. Subjects were instructed to focus on the target stimuli and silently count how many times they flashed while avoiding eye blinking. Targets were a random sequence of 18 commands, each command is randomly repeated twice. 

In the online test phase the same protocol was kept with the only difference being the number of sequence set to one i.e. single trial, an additional 1 second after the last flash was added for processing and feedback. The feedback was presented for 500 ms as a blue square on top of the icon selected by the classifier in case of wrong decision, otherwise a green square highlighted the correct target indicating successful detection, and thus the selection time for a single command is 2.49 seconds. During each copy-spelling session subjects selected 18 icons (with the exception of S3 who spelled 27 commands), the order of the desired sequence was random. Figure 2 illustrates the experimental protocol.	

After the session end, subjects filled experiment questionnaire forms. The speller is implemented in our C++ Qt framework (\url{https://www.qt.io}) based open source stimulation platform StimUSTO (\url{https://github.com/okbalefthanded/StimUSTO}). For online processing our python based open source pyLpov (\url{https://github.com/okbalefthanded/pyLpov}) library was used. The real-time platform OpenVIBE \cite{38} controls the entire process and communicates with the stimulation platform using TCP/IP protocol for event marker tagging and UDP protocol for feedback.

\subsection{Data acquisition}

\begin{figure*}[t]
\centering
\includegraphics[width=\columnwidth]{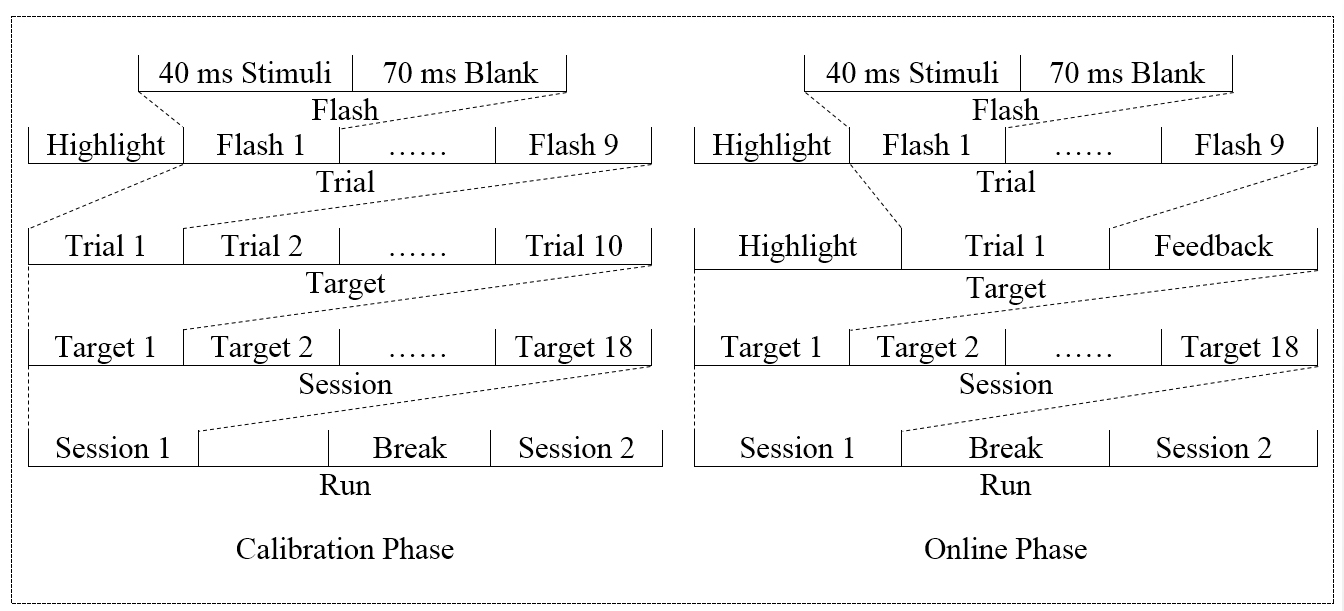}  
  \caption{Experimental protocol for the calibration and online phases.}
  \label{fig:2}
\end{figure*}

\begin{figure}[h!]
\centering
\centerline{\includegraphics[width=.5\columnwidth]{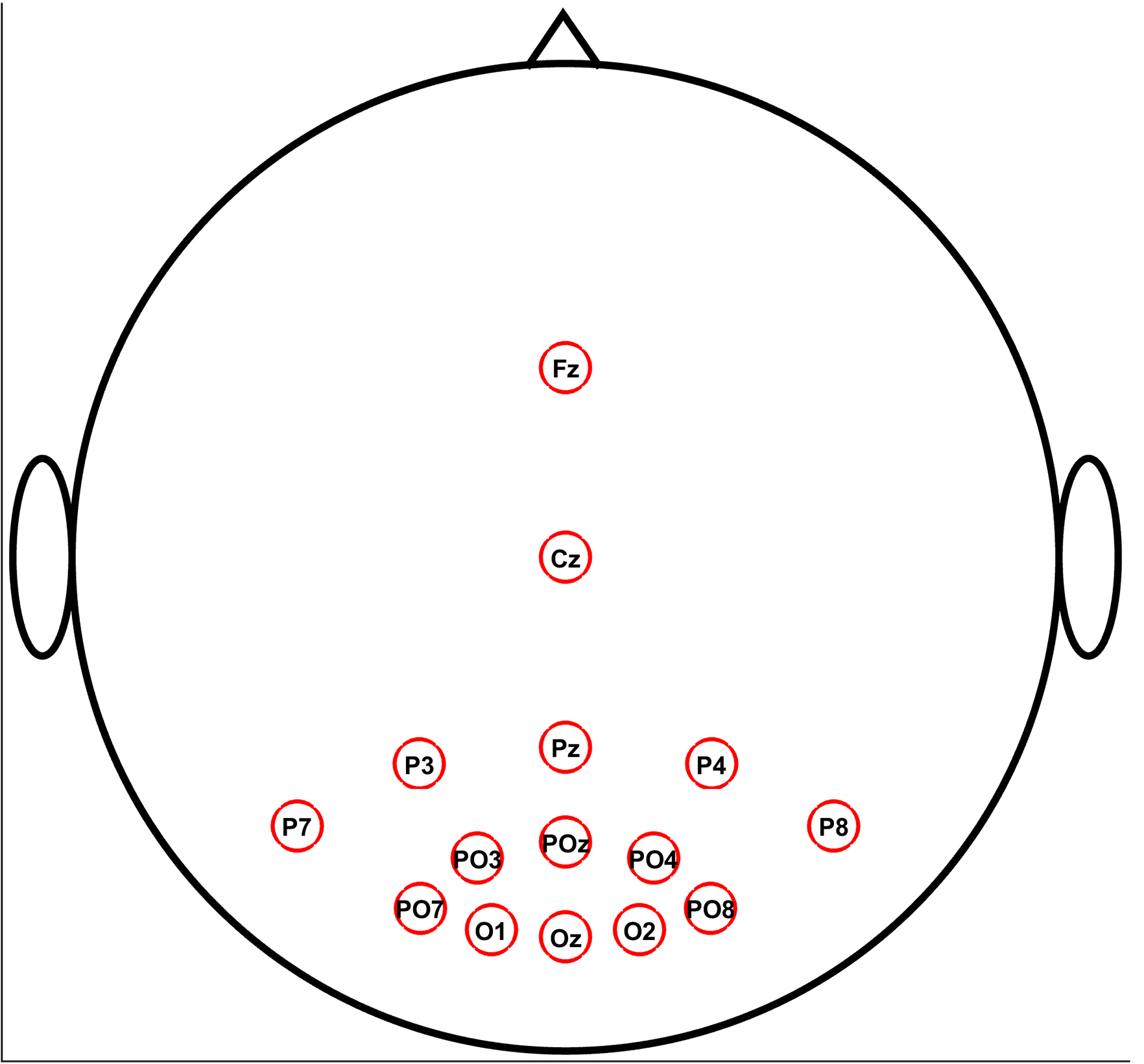}}  
  \caption{Electrodes montage.}
  \label{fig:3}
\end{figure}

EEG signals were recorded from 15 g.Sahara dry electrodes with the gUSBamp amplifier, at a sampling rate of 512Hz. A 50Hz notch filter to remove AC artifacts and a Butterworth bandpass filter of 0.1-60Hz with order 4 were applied. Ground electrode was mounted on the left mastoid and the reference was set to the right mastoid. The following electrodes were mounted on the EEG cap following the 10-20 international system: PO7, P3, P7, Fz, Cz, Pz, POz, PO3, O1, Oz, O2, P4, P8, PO4, PO8 (see Figure 3) with impedance kept below 5k\textOmega. 

\subsection{Data Analysis}
\subsubsection{Offline and online analysis}
To observe ERP components, the data were first band pass filtered between 1-10 Hz with a 2nd order zero-phase Butterworth filter, then overlapping epochs of 800 ms post-stimulus following event markers timing were extracted after a 200ms pre-stimulus mean subtraction baseline correction, thus for each calibration session 1620 epochs (9 flashes x 10 repetitions x 18 commands) were segmented from the continuous signal to form an epochs tensor of 410 x 15 x 1620 in dimension (samples, channels, trials). No artifact correction or rejection operation was applied. BBCI toolbox \cite{39} was used for the offline analysis and visualizations. 
To evaluate the ERPs’ discriminative information, signed r2-values are computed based on the pointwise biserial correlation coefficient (r²-value) \cite{40}:
\boldmath
\begin{equation}
\operatorname{signed} r^2(x)=\operatorname{sign}(x) \cdot\left(\frac{\sqrt{N_1 N_2}}{N_1+N_2} \cdot \frac{\operatorname{mean}\left\{X_1\right\}-\operatorname{mean}\left\{X_2\right\}}{\operatorname{std}\{X\}}\right)^2
\label{rsquare}
\end{equation}

where $X_{1}$ and $X_{2}$  are the trials belonging to class 1 (target) and class 2 (non-target) respectively, and X is all the trials. $N_{1}$ and $N_{2}$ are the number of trials in each class.

For Online classification, both the calibration and test data were baseline corrected by mean amplitude subtraction in the -200 to 0 ms pre-stimulus interval from every event marker. Since the relevant ERPs appear in the interval of 100ms to 500 ms post-stimulus and the short ISI of 70 ms employed has higher risk of overlapping ERPs, the epoch duration is shortened to 400 ms starting from 100 ms post-stimulus. This selection also reduces the dimensionality of the signal allowing for faster processing.  
Another difference from the offline analysis is the filter frequency band, as higher power in the delta (0.1-4 Hz) and theta (4-8 Hz) frequency bands was observed in the dry electrodes used in this study \cite{24} , we opted for a frequency band filter of 5-12 Hz.

The superiority of EEGNet \cite{28} against other state-of-the art feature extraction and classification methods and its disposition to handle limited training data favored its choice. In this study we employ the default EEGNet 8-2 configuration for ERP data. Except for a z-score normalization, to ensure a full end-to-end training approach no further pre-processing or data augmentation operations were applied before model training. 

For each subject, a specific model was trained on all data available for that subject without correction for class imbalance.  The model minimizes the binary cross entropy and was trained for 500 epochs with a batch size of 64 using AdamW \cite{41} optimizer with a fixed learning rate of 1e-3 and weight decay of 1e-4.  The model training was achieved on a Google Colaboratory NVIDIA Tesla T4 GPU in PyTorch \cite{42} using the Aawedha deep learning toolbox (\url{https://github.com/okbalefthanded/aawedha}). After model training, the online classification was run on the same laptop (4GB RAM, Intel i7-2670QM CPU @ 2.20GHz) with the stimulation presentation.

The decision process to select a command \textbf{\textit{C}} is a two-stage operation, at first the model output a score of prediction \textbf{\textit{S}} for each epoch being the target class, S can be a probability for models with outputs that can be directly interpreted as a confidence level $\textbf{\textit{S}} \in \lbrack0,1\rbrack$ or the value of the decision function $(S \in \mathbb{R})$ for other types of models. Then the icon with the maximum score is chosen as the command:

\begin{equation}
C = \argmax_i S(X_i) \indent{i = 1...9}
\label{commad}
\end{equation}
where $X_i$ is an EEG epoch of 1x15x205 (1, Channels, Samples) in dimension. We assess the system performance by reporting the following metrics on online test data: balanced accuracy (BA) \cite{43}, Area Under the Curve (AUC) of the Receiver Operating Characteristic (ROC) \cite{44}, information transfer rate (ITR) \cite{45}  and correct command detection rate defined as: 
\begin{equation}
BA = \frac{1}{2}(\frac{TP}{TP+FN} + \frac{TN}{TN+FP})
\label{eq:03}
\end{equation}
where  TP, TN, FP and FN are the number of true positive trials, true negative trials, false positive trials and false negative trials, respectively of the binary classification. And:
\begin{equation}
ITR = \{\log_2 N + p\log_2 p + (1-p)\log_2 \frac{1-p}{N-1}\}\frac{60}{T}
\label{eq:04}
\end{equation}
where \textit{N} is the number of commands,  \textit{p} is the classifier accuracy and T is the time required for the making of a selection. Here the correct command detection rate is the accuracy of commands classification. 

\subsubsection{Models comparison}
To further examine the system’s performance, we compared conventional and state-of-the art feature extraction and classification methods against different CNNs developed for Brain-Computer Interface data classification. As a baseline, the classical downsampling, moving average and channels concatenation method \cite{46} is tested with two linear discriminant analysis (LDA) classifiers, the stepwise-LDA (SWLDA) and the regularized LDA with covariance shrinkage (LDA-shrinkage) \cite{40}. The same factor for moving average window and downsampling is set as 12. The widely applied approach based on downsampling, windsorizing and Bayesian LDA (BLDA) \cite{47} was also tested with its default parameters: downsampling factor of 12, replacing the values below the 10th percentile or above the 90th percentile by the 10th percentile  or 90th  percentile respectively. 

The higher performant approaches based on Riemannian Geometry (RG) was also tested, specifically the Kaggle BCI challenge winning entry (available at \url{http://github.com/alexandrebarachant/bci-challenge-ner-2015}), which combines xDAWN spatial filtering \cite{48}, with the projection of covariance matrices onto the tangent space \cite{49} followed by Elastic Net regression for classification. 

Besides the exclusion of meta-features, channel selection and ensemble of classifiers from the original processing pipeline, we kept the same configuration of parameters as this approach (we will refer to it as xDAWN+TS+EN). We added a different pipeline by changing the metric of tangent space projection from log-euclidean to Riemann distance, z-score feature normalization in place of L1 normalization and linear Support Vector Machines (SVM) for classification (we will refer to it as xDAWN+TS+SVM). Baseline methods are implemented following the scikit-learn API \cite{50,51}, the implementation is available at (\url{https://github.com/okbalefthanded/BCI-Baseline}). The Riemann geometry approaches are implemented using the python pyRiemann package (\url{https://github.com/pyRiemann/pyRiemann}).

Restricted by the small amount of data available in the single subject training scheme adopted in this study, neural networks with large number of trainable parameters are easily prone to overfitting and weaker generalization capacity, furthermore, the authors in \cite{52} showed that small and medium size models perform better or are on par with larger models, therefore we keep the three best performing models from that study: SepConv1D \cite{52}, EEGNet \cite{28}, and DeepConvNet \cite{27} in this comparison, with two more models added : EEGTCNet \cite{53} and EEG-Inception \cite{54}. Here we briefly describe the models, for more details on the architectures see appendix A. 

SepConv1D serves as a simple baseline for CNNs, it consists of only two layers, a 1D Separable convolution with a Hyperbolic Tangent (tanh) as an activation function, followed by a single sigmoid activated output neuron. It uses kernels of length 16 x C with stride 8. (see Appendix A.1)

EEGNet is a two-blocks compact CNN, in the first block the first layer acts as temporal filters by employing typical 2D convolution of (1 x half sampling rate) length, while the second layer serves as spatial filters using 2D Depthwise convolution of (C x 1) length to reduce the number of trainable parameters, the second block uses 2D Separable convolution to reduce the number of trainable parameters and decouple the relationship within and across feature maps. Finally, the classification is handled by a Softmax dense layer in case of categorical labels, here we use the sigmoid activation function for a single output. In the two blocks, after each convolution Batch normalization \cite{55} is applied before an ELU \cite{56} activation, then to reduce the dimension of feature maps a 2D average pool is applied.(see Appendix A.2)
	
EEGTCNet is an extension to EEGNet introduced in the context of Motor-Imagery classification to construct an accurate model while keeping the model size, inference time and memory print small for embedded systems use. To do so, layers of temporal convolutional networks (TCN) \cite{57} are stacked on top of the last convolution block in EEGNet. A TCN block consists of layers of dilated causal convolutions, followed by batch normalization and a dropout in between the layers, lastly, a skip connection adds the input to the output feature map creating a residual block. (see Appendix A.3)	
	
EEG-Inception adapts the inception \cite{58} module to ERP classification to learn features at different scales, the model consists of two inception modules followed by an output module. In the first inception module, two layers of three parallel convolution blocks are stacked together, each block is a sequence of convolution, batch normalization, activation and dropout operations. The difference between the blocks resides in the type of convolution and the kernels size, the first layer uses 2D convolutions while the second layer uses 2D Depthwise convolution. The second inception module is three parallel convolutions blocks with different scales. The output module has two convolution blocks separated by Average pooling. The three modules are connected by two concatenation and average pooling layers. The last layer is the typical softmax classification layer. (see Appendix A.5)

\begin{table}
\centering
\begin{tabular}{|l|l|l|} 
\hline
\begin{tabular}[c]{@{}l@{}}\\  Evaluation\end{tabular} & Online          & Comparison       \\ 
\hline
Optimizer                                              & AdamW           & AdamW            \\ 
\hline
Learning
  rate                                        & 1e-3            & 1e-3             \\ 
\hline
Weight
  decay                                         & 1e-4            & 1e-4             \\ 
\hline
Optimizer
  momentum                                   & ~$\alpha$ = (0.9, 0.999) & ~$\beta$ = (0.9, 0.999)  \\ 
\hline
Epsilon                                                & 1e-7            & 1e-7             \\ 
\hline
Batch
  size                                           & 64              & 64               \\ 
\hline
Training
  epochs                                      & 500             & 250              \\
\hline
\end{tabular}
\caption{Convolutional neural networks training settings per evaluation.}
\end{table}

\begin{figure}[t]
    \centering
    \includegraphics[width=\columnwidth]{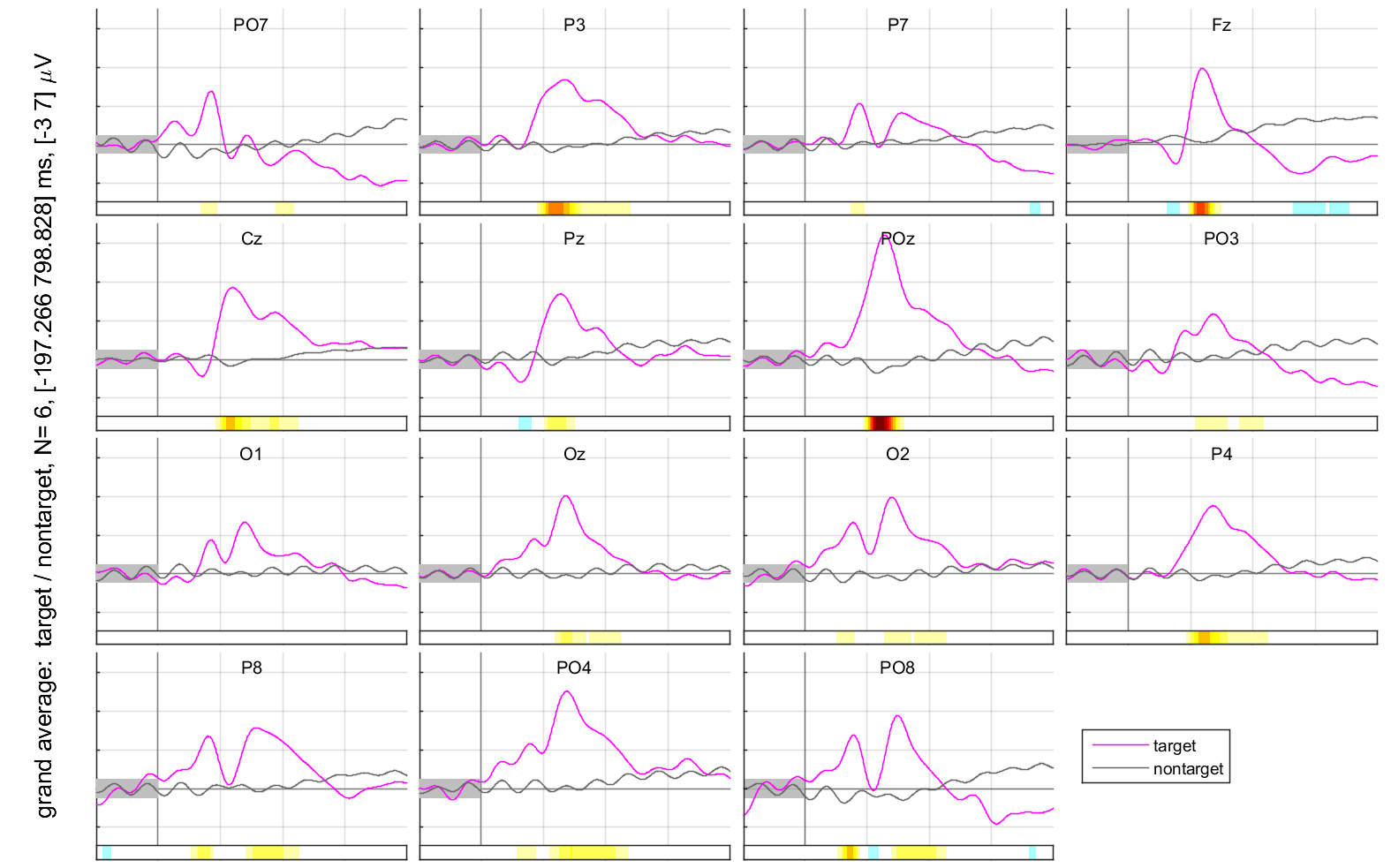}  
    \caption {Grand average ERP waveforms, the target class is presented in magenta, non-target class in gray. The bar below channel plots depicts the time intervals with significant r2-values as a colorbar, the yellow to red color indicate amplitude from target epochs are higher, whereas blue color refers to target epochs amplitudes lesser than non-target epochs.}
    \label{fig:4}
\end{figure}

DeepConvNet is the largest among our selection, with the exception of the first block containing two convolutions serving as temporal filters followed by a spatial filters, the remaining three blocks followed by a softmax classification layer consist of convolution, batch normalization, activation, max pooling and dropout. The number of kernels is increasing from block to block. ELU activation function is used throughout the network.	(see Appendix A.8)

Except for DeepConvNet, in which we followed the implementation of the EEGNet paper authors, the rest of the models are re-implemented following their original implementation with the default parameters and options. The models were trained with the same preprocessing and configuration as in the online processing, only the number of epochs was reduced to 250. The training procedure configurations for the online and comparison evaluations are summarized in Table 3. 
	
We run the training and testing for each model on each subject 30 times and report the average on the whole data set (30 iterations, 6 subjects, and 2 sessions) of the same performance metrics as the online experiment except for balanced accuracy that was omitted from the evaluation due to ElasticNet being a regressor and we followed the original approach. Since the test data are independent of the training data, no cross-validation or train-validation split were performed.

\subsubsection{Models complexity}
Apart from the control performance, we consider the difference between deep models in terms of complexity. We assess the following elements: Number of trainable parameters, training time, inference time, multiply-accumulate (MAC) operations per inference count, and ratio of parameters. The training time is reported as the mean of training time for each model from the model comparison evaluation measured in minutes. The inference time was calculated on the online test machine as the mean of 10 iterations of the inference of a single trial. MAC is reported by thousand operations, and the ratio of parameters is measured the number of training samples divided by number of trainable parameters.

\subsubsection{Statistical analysis}
For testing the significance of differences between classical approaches and neural networks across subjects in the models comparison evaluation, nonparametric tests were adopted, a Friedman test followed by a Wilcoxon signed-rank test were applied on the three performance metrics AUC, command detection rate and ITR. For the experiments subjective reports a Wilcoxon signed-rank test between calibration and online surveys was applied. In all tests the significance level was set to the value 0.05.

\section{Results}
\subsection{Offline Analysis}
Figure 4. Depicts the target and non-target ERP waveforms averaged across all 6 subjects over the 15 electrodes with r2-values bar below the channel plots highlighting time intervals where significant difference between the two conditions occurs. A clear distinction between the two conditions is seen on each channel, the most prominent component is apparently peaking in the P300 interval of 200-400 ms after stimulation onset specifically on central, parietal and parietal-occipital areas, in line with previous studies on face stimulation \cite{15, 17}, indicating high attention to stimulus.  The face specific N170 is weakly modulated at PO8 with an amplitude of -0.0947µV and latency of 219.8852 ms at PO8 as it is seen on the lateral parietal areas, in line with \cite{59} that reported face inversion coupled with color caused a decline in N170 modulation. A pronounced N100 at central frontal and parietal channels is apparent.

\begin{figure}[h!]
    \centerline{\includegraphics[width=\columnwidth]{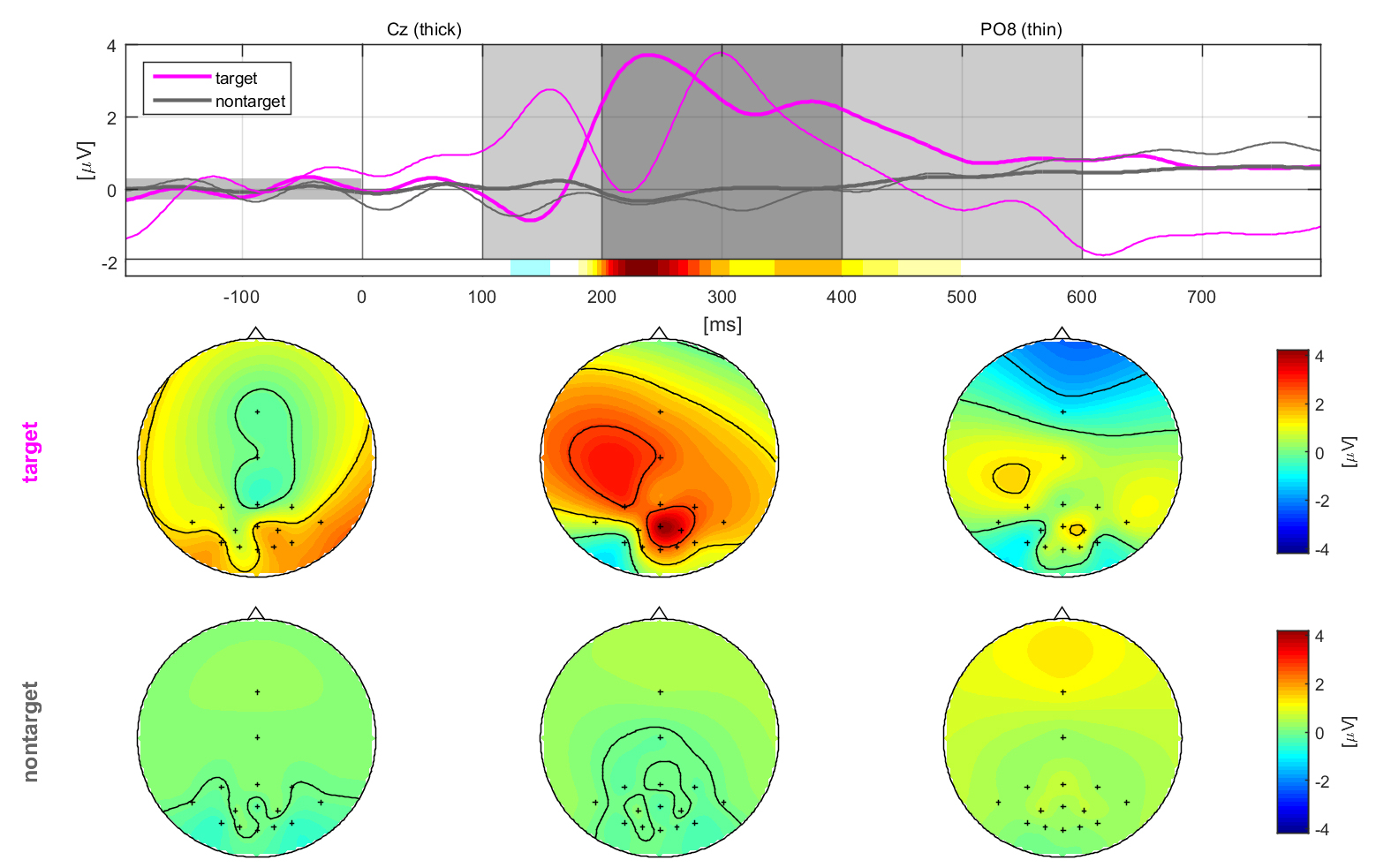}}  
    \centerline{\includegraphics[width=\columnwidth]{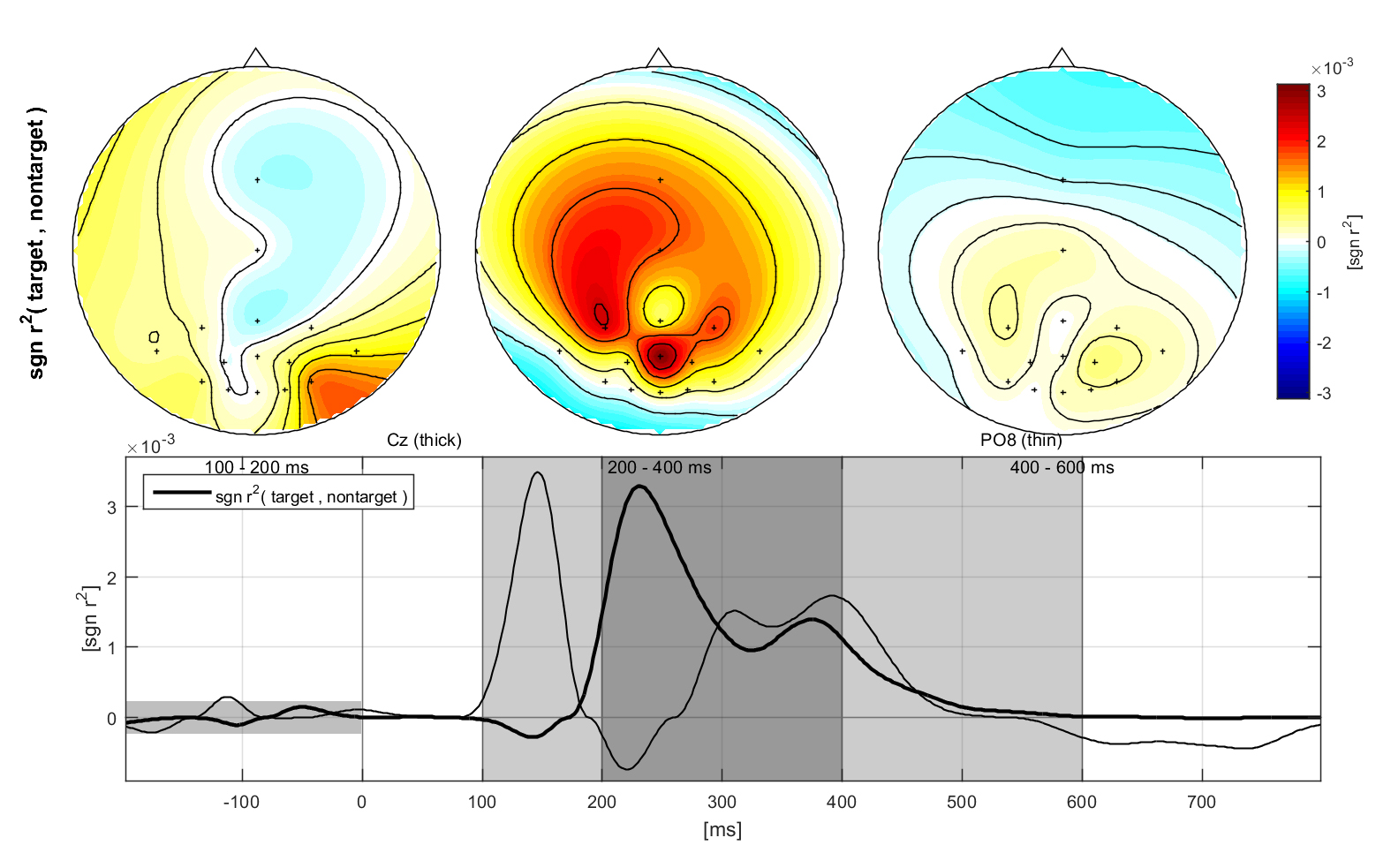}}  
    \caption {Top: grand average ERP waveform for target class in magenta (Cz thick, PO8 thin), non-target class in gray. Middle: scalp topographic map of selected intervals of ERP waveforms and signed r2. Bottom: temporal evolution of signed r2 in Cz (thick) and PO8 (thin).}
    \label{fig:5}
\end{figure}

As can be seen in Figure 4, the P100 at lateral parietal and occipital sites (P7, P8,O1, O2, P8, PO8) is larger in the interval of 80-160ms post-stimulus, these findings corroborate the results reported in \cite{59} on the effect of face inversion on P100 amplitude and latency. They showed significant increase in amplitude for inverted colored face over upright natural face in the same regions.
The VPP is evoked in the fronto-central sites at around 200 ms in good agreement with established literature \cite{17, 20}. Lastly, the familiar face sensitive N400 ERP is evoked only in Fz and slightly in Pz in the interval of 400-600 ms post-stimulus.
Peak amplitudes and latencies were determined from the grand average for each component in their related interval at the channel with the highest amplitude, we selected the three intervals where the ERPs are more distinct, first 100-200ms post-stimulus for early ERPs N100, P100 and N170/N200, located in Pz, P8 and PO8 respectively. 200-400 ms post-stimulus for VPP in Fz and P300 in POz. The N400 was fixed in 400-600 ms post-stimulus at Fz. The peaks and latencies of the grand average values are presented in Table 4. 
N100 had latency of 128 ms, P100 had latency at 157.51 ms, this latency is larger than reported in similar studies \cite{17, 19, 20}. VPP had a latency of 237 ms in line with the literature. P300 had a latency of 256.93 ms in line with \cite{17}, N400 had a latency of 551.27 ms in line with the literature \cite{17, 20, 21}.
The grand average of selected channels Cz and PO8 alongside topographic plots of grand average amplitude and signed r2-values with the three selected analysis intervals are illustrated in Figure 5 Top. In the first interval of 100-200 ms post-stimulus negative amplitudes activity is spread over central and frontal areas corresponding to N200 modulation, while a high positive amplitudes activity concentrated more in the right hemisphere is present, which corresponds to P100 modulation in line with literature \cite{59} associating this time frame with processing of face. The 200-400 ms is dominated by positive amplitudes distributed over most electrodes with higher activity in frontal, central and central-parietal regions, corresponding to the VPP/P300 responses. The last interval of 400-600 ms post-stimulus is marked by the decrease of activity all over the scalp, with higher negative amplitudes in frontal and late occipital-parietal electrodes hinting at N400 modulation.

In Figure 5 Bottom, The apparent  spatio-temporal patterns support our selection of a shorter time window for epoch segmentation in the online analysis, most of the discriminative activity in both amplitude polarities is restricted in the interval 100-500 ms post stimulus.

\subsection{Online Analysis}
Table 5 indicates the participants’ online single trial performance measured by balanced accuracy, AUC, information transfer rate and command detection rate. We report the average of the two online sessions for each subject. We observe a division of results into two categories, on one hand, superior performance passing 80\% in command detection rate and 45 bit/min, with two subjects above 90\% and 60 bit/min among them a participant almost reaching the 100\% bar. However, on the other hand, weak performance are exhibited by the remaining subjects although, clear and distinct ERPs were evoked in the calibration data. This deterioration is mainly the effect of high data shift present in EEG data which is enforced in dry electrodes, as it was demonstrated in studies comparing the same equipment we used against other type of wet electrodes \cite{24, 23}. The dry electrodes ranked the last in all evaluations in terms of signal-to-noise ratio, ERP amplitude and latency, and classification accuracy.  
To showcase the inter-session shift between calibration and testing data, we plot the grand average ERP waveforms for both calibration sessions and the online sessions, illustrated in Figure 6. The same analysis configuration as the offline analysis was employed. The amplitudes reveal an evident drop in values in many channels, the most severe decline is noticed in lateral parietal and parieto-occipital sites. On the contrary, frontal and central electrodes manifested a different change, where the online sessions had higher amplitude, this is mainly due to the higher attention directed to the single-trial stimulation in the online experiment, and reduced repetition effect.

\begin{table}
\centering
\begin{tabular}{|l|p{2cm}|l|p{2cm}|p{2cm}|} 
\hline
ERP component & Channel & Amplitude (µV) & Latency (ms)  \\ 
\hline
N100          & Pz      & -1.18          & 128.27        \\ 
\hline
P100          & P8      & 2.76           & 157.51        \\ 
\hline
VPP           & Fz      & ~3.96          & 237.43        \\ 
\hline
P300          & POz     & 6.41           & 256.93        \\ 
\hline
N400          & Fz      & -1.49          & 551.27        \\
\hline
\end{tabular}
\caption{Average peak amplitudes and latencies of ERP components. }
\end{table}

\begin{table}[t]
\centering
\begin{tabular}{|l|p{2cm}|l|p{2cm}|p{2cm}|} 
\hline
Participant & Balanced
  Accuracy (\%) & AUC           & Command
  detection rate (\%) & ITR (bit/min)   \\ 
\hline
S1          & 86.98                    & 0.98          & 86.11                         & 52.45           \\ 
\hline
S2          & \textbf{97.92}           & \textbf{0.99} & \textbf{97.22}                & \textbf{70.65}  \\ 
\hline
S3          & 86.23                    & 0.96          & 81.47                         & 46.99           \\ 
\hline
S4          & 80.73                    & 0.89          & 61.10                         & 25.27           \\ 
\hline
S5          & \textbf{90.80}           & \textbf{0.99} & \textbf{91.66}                & \textbf{62.53}  \\ 
\hline
S6          & 69.97                    & 0.86          & 55.56                         & 20.60           \\ 
\hline
Average     & \textbf{85.44}           & \textbf{0.95} & \textbf{78.86}                & \textbf{46.41}  \\ 
\hline
STD         & 8.655                    & 0.048         & 15.38                         & 18.25           \\
\hline
\end{tabular}
\caption{Online performance for each subject with EEGNet averaged across 2 sessions. Subjects with high performance and average values are highlighted in bold.}
\end{table}

\begin{figure}
\centering
\centerline{\includegraphics[width=\columnwidth]{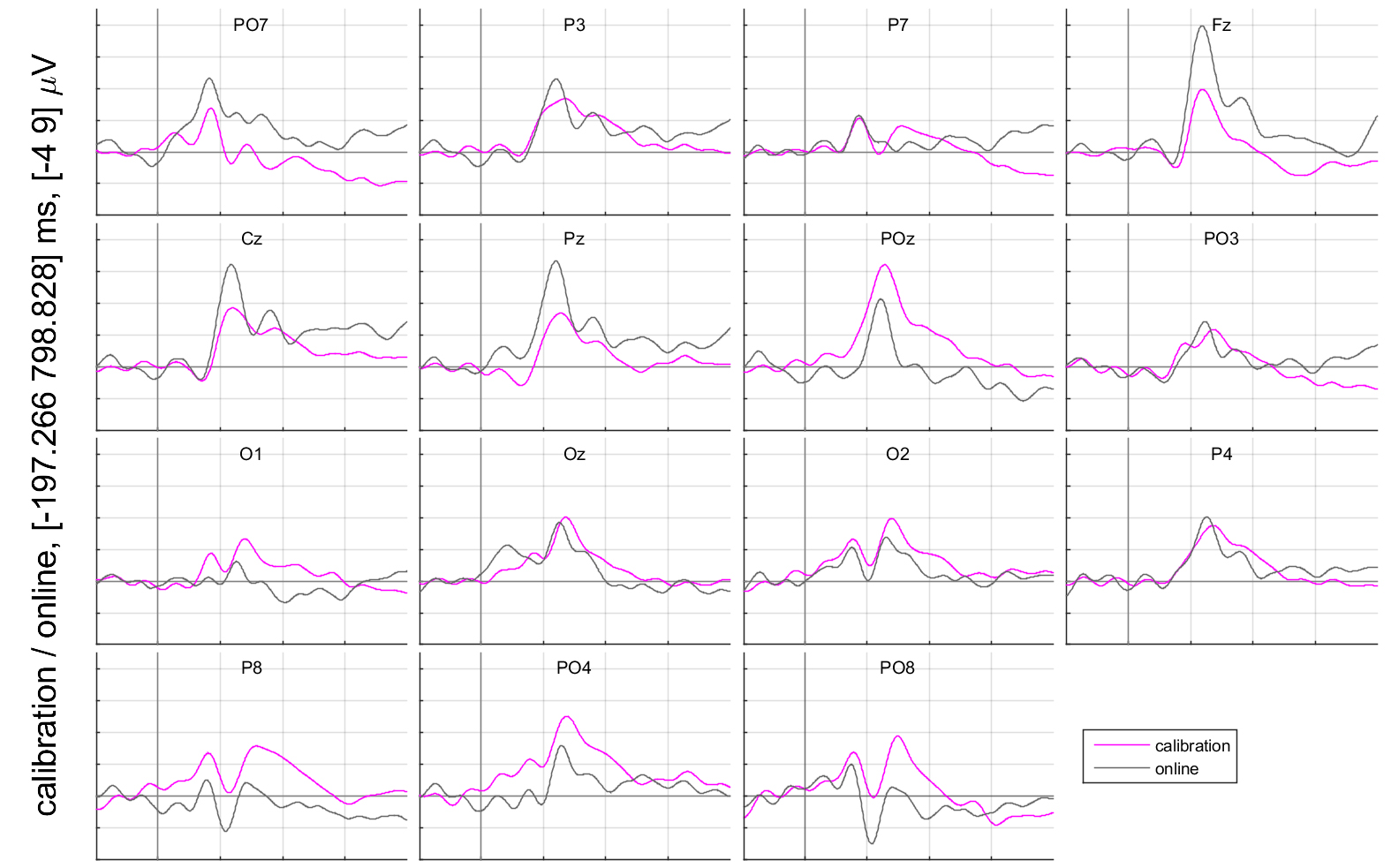}}  
  \caption {Calibration versus Online sessions’ grand average ERP waveforms.}
 \label{fig:6}
\end{figure}

\subsection{Models Comparison}
The results on models comparison are presented in Table 6. The values are reported by the mean and standard deviation over all repetitions across all subjects. The Friedman test showed a significant difference between the models, in AUC ($\chi^2=22.10, p=0.0085 < 0.01$) and ITR ($\chi^2=17.27, p=0.044 < 0.05$), however the command detection rate difference was marginally significant ($\chi^2=16.0143, p= 0.066 > 0.05$).  
	
Regarding the Wilcoxon signed rank test, AUC showed significant ($p  < 0.05$) difference for pairs of SepConv1D against all LDA based models, SepConv1D against EEGNet and EEGTCNet, and lastly DeepConvNet was outperformed by Sh-LDA, EEGNet and EEGTCNet. 

The command detection rate difference was significant ($p < 0.05$) only between SepConv1D against Sh-LDA, EEGNet and EEGTCNet. The ITR difference was significant ($p < 0.05$) between SepConv1D against Sh-LDA, EEGNet and EEGTCNet, and EEGTCNet outperformed EEGTNet.
Marginal significant difference ($p = 0.0625$) in AUC was noticed for two models, first, between EEGInception and Sh-LDA, EEGNet and EEGTCNet, and between DeepConvNet and SWLDA. Command detection rate registered marginal significance ($p = 0.0625$) between SepConv1D and Sh-LDA, EEGInception. EEGTCNet had the same effect with SWLDA, TS+xDAWN+EN, and the 3 other deep networks models. The same effect was seen in ITR for EEGTCNet, also between EEGInception and SepConv1D.
	
In contrast to \cite{52}, where only neural networks based methods were tested, SepConv1D was outperformed by most models. Among its peers, SepConv1D was the least performing model, followed by the larger DeepConvNet and wider EEGInception, while SepConv1D only layer could not learn richer representation, the large and wide models suffered most from overfitting as the training samples per parameters were too small. 
The Compact architectures EEGNet and its extension EEGTCNet were the best performing in all metrics, with EEGTCNet surpassing all the models. The inconclusive results between the different classes of models is due to the variability within subjects, as some models perform worse for certain subjects, Figure 7 shows the best model’s command detection rate from each category for the best performing subjects S1, S2, S3, and S5.	

\begin{table*}[t]
\centering
\begin{tabular}{|l|l|l|l|} 
\hline
Method       & AUC                      & Command
  detection rate (\%)       & ITR (bit/min)                        \\ 
\hline
Sh-LDA       & 0.9377
  ± 0.0511        & 76.3889
  ± 18.1974                 & 44.1589
  ± 20.4009                  \\ 
\hline
SWLDA        & 0.9328 ± 0.05811         & 74.0741 ± 18.4722                   & ~41.4716 ± 20.307                    \\ 
\hline
BLDA         & 0.9332 ± 0.06188         & 74.2284 ± 19.1663                   & 41.525 ± 20.0936                     \\ 
\hline
Xdawn+TS+EN  & 0.92483 ± 0.0648         & 73.7654 ± 18.4928                   & 40.8734 ±18.9925                     \\ 
\hline
Xdawn+TS+SVM & 0.9291
  ± 0.05068       & 74.5370 ± 15.91                     & 41.0804 ±16.4635                     \\ 
\hline
SepConv1D    & 0.9183
  ± 0.0648        & 71.6101
  ± 17.858                  & 38.5133
  ± 17.3904                  \\ 
\hline
EEGNet       & 0.938
  ± 0.054          & 77.7932
  ± 15.9312                 & 45.2213
  ± 17.8603                  \\ 
\hline
EEGTCNet     & \textbf{0.9407 ± 0.0522} & \textbf{80.0514 }± \textbf{15.4381} & \textbf{47.8746 }± \textbf{18.3284}  \\ 
\hline
EEGInception & 0.9196
  ± 0.0565        & 75.6944 ± 15.8606                   & 42.5947 ± 17.0395                    \\ 
\hline
DeepConvNet  & 0.9199
  ± 0.05229       & 74.8508
  ± 14.4267                 & 41.4923
  ± 16.1515                  \\
\hline
\end{tabular}
\caption{Conventional feature extraction and classification approach comparison against convolutional neural networks. The average of each metric in a 10 repetition evaluation across all subjects with standard deviation.}
\end{table*}

\subsection{Models complexity}
Table 7 summarizes the models properties, the parameters per samples ratio favors EEGTCNet, and it is the least effected by the data sample size. Moreover, all models displayed reasonable training times that can run in parallel while the subject rests before performing online sessions without taking notice of any delay. It is important to recall that the training was executed on remote GPU and the inference on a local CPU. The compact architectures EEGNet and EEGTCNet provide the best trade-off between model complexity and performance, with latter outperforming its base counterpart.

\begin{figure}
  \centering
  \includegraphics[width=\columnwidth]{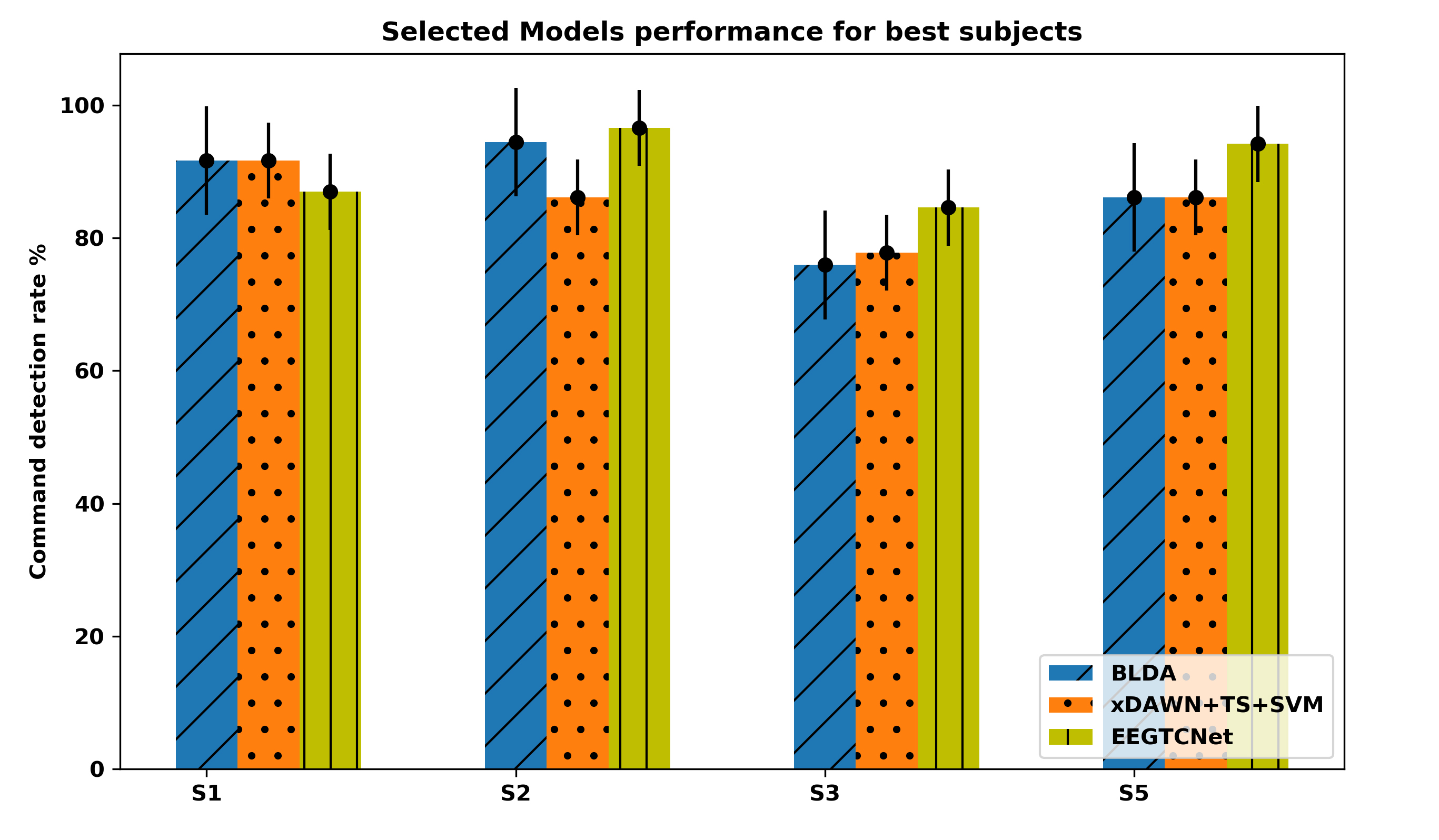}
  \caption{Models command detection rate for best performing subjects. Error bars denote 2 standard errors of the mean.}
  \label{fig:7}
\end{figure}

\begin{table}
\centering
\begin{tabular}{|l|p{2cm}|l|l|p{2cm}|p{1cm}|} 
\hline
Model        & Trainable
  parameters & Ratio & \begin{tabular}[c]{@{}l@{}}Train time \\(mm:ss:msec ± msec)\end{tabular} & Inference
  time (msec) & MAC (K)  \\ 
\hline
SepConv1D    & 412                    & 3.93  & 00:49.9
  ± 09.35                                                        & 6.84 ± 0.006            & 7.6      \\ 
\hline
EEGNet       & 1,185                  & 1.37  & 01:02.2
  ± 12.65                                                        & 13.12 ± 0.04            & 978.3    \\ 
\hline
EEGTCNet     & 3,945                  & 0.41  & 01:33.4
  ± 19.7                                                         & 20.12 ±0.006            & 982.36   \\ 
\hline
EEGInception & 15,273                 & 0.11  & 01:47.5
  ± 21.7                                                         & 66.93±0.035             & 3856     \\ 
\hline
DeepConvNet  & 143,301                & 0.01  & 05:01.1±
  55.98                                                         & 39.53 ±0.008            & 5834     \\
\hline
\end{tabular}
\caption{Tested Convolutional neural networks properties. mm = minutes, ss = seconds, msec = milliseconds, K = thousand.}
\end{table}

\begin{figure}
\centering
\centerline{\includegraphics[width=\columnwidth]{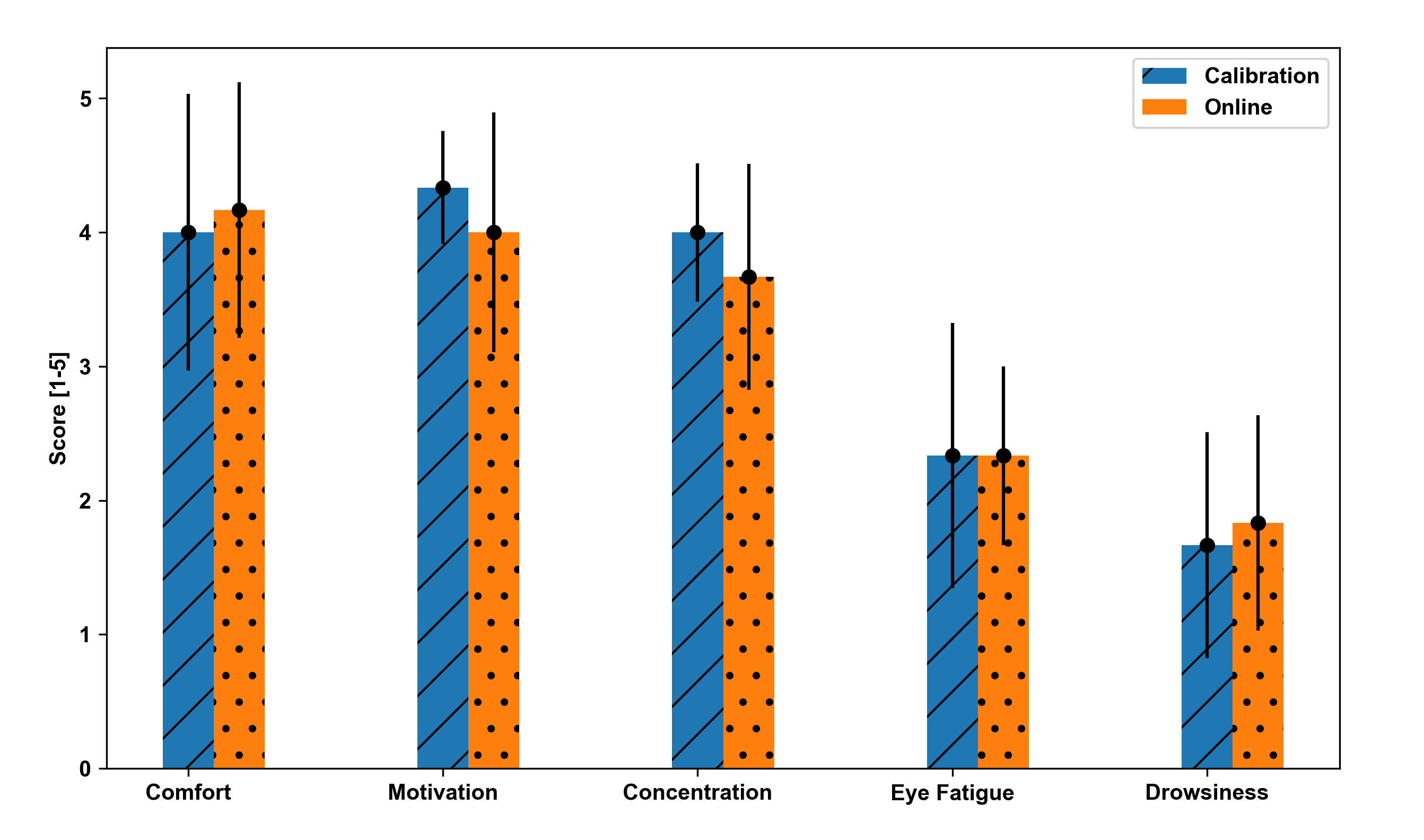}}  
\caption {Questionnaire responses in comfort, motivation, concentration, eye fatigue and drowsiness. Means averaged across subjects for calibration and online sessions, respectively. Error bars denote 2 standard error of means. 1 = low, 5 = High.}
\label{fig:8}
\end{figure}

\begin{figure}
\centering
\centerline{\includegraphics[width=\columnwidth]{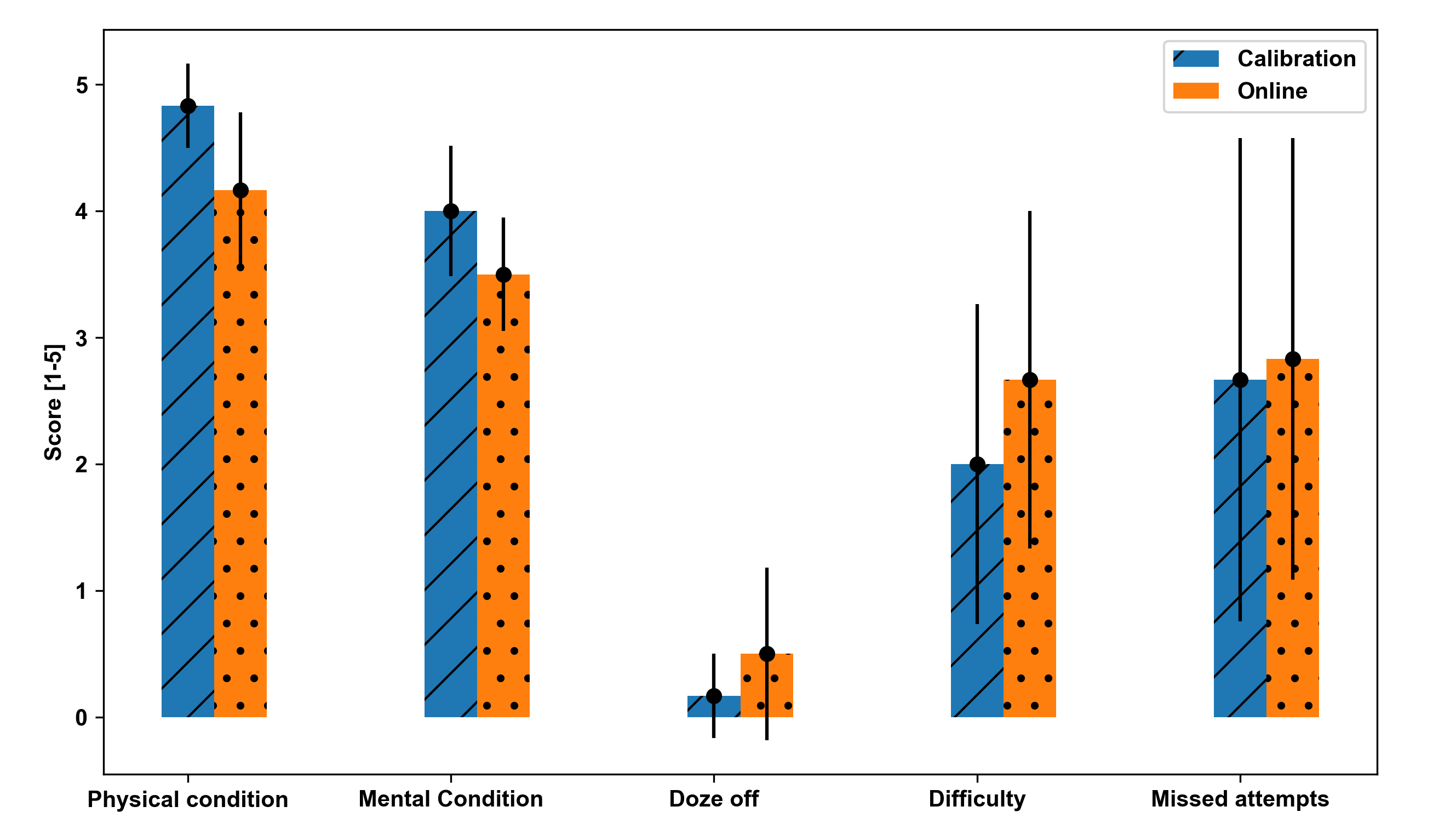}}  
  \caption{Questionnaire responses in physical condition, mental condition, doze off, difficulty, and missed attempts. Means averaged across subjects for calibration and online sessions, respectively. Error bars denote 2 standard error of means. 1 = low, 5 = High.}
\label{fig:9}
\end{figure}

\subsection{Subjective report}
Figure 8 and 9 show the mean value responses from the subjective reports filled by subjects after the completion of the calibration and online sessions, respectively. Although slightly different, the scores indicate that the calibration session was more motivating, less demanding to the physical and mental condition. On the contrary the online session was more difficult to achieve and required more concentration, this is mainly due to the rapid stimulation. Both sessions caused equal mild eye fatigue. The statistical test failed to show any significant difference between sessions ($p >0.05$).

\section{Discussion}
Most works on face stimulus based ERP BCI, mainly focused on optimizing a single parameter to improve the overall system performance. The combination of different elements to construct an optimized system remains a daunting challenge. Therefore, we aimed at merging different improvements in each part of the system. First, to elicit strong and robust ERP components, we combined the configuration of orientation and color in a face stimulus, second, to speed up the throughput while keeping the implementation requirements simpler, we fixed the round of a single command production to 1 second (independent of cue and feedback), which imposes a reduction in stimulus exposure time and ISI. The quick setup and ease of use of dry electrodes comes with the drawback of lower SNR which negatively impacts the system’s performance, thus we employed convolutional neural networks for classification, known to possess a good robustness properties against data shift \cite{60}. Another major contribution of our work is the demonstration of successful deployment of deep neural networks in a regular laptop. For long after their huge success in computer vision and other domains, deep neural networks were neglected by the BCI research community, with the main argument for this skepticism driven by the scarcity of EEG data required for training large networks and the preference for simpler linear models to avoid the risk of overfitting to irrelevant signal components. Nevertheless, on small data size we can still manage to train well designed architectures.

The weak modulation of N170 might in fact be attributed to additional factors. Firstly, the experiment short stimulation and ISI. As described before in studies on flashing paradigm \cite{61} and motion-onset paradigm \cite{62}, the early negative components N100 \cite{62} and N200 \cite{61} are overshadowed by the P100 positive peak in short ISI settings. Secondly, the shape of waveforms at P7, P8, PO7 and PO8 is similar to conventional ERPs at those sites, but had amplitude around zero, which might be an effect of motion artefacts polluting the signal , as reported in \cite{24} for the same electrodes used in this study.
Also, the absence of N400 in central sites suggests an unfamiliarity of some subjects with the face used as stimulus, the observation is in line with \cite{19}. The fail to evoke the N400 was credited to unfamiliarity with the face stimulus, furthermore, in the BCI based on inverted face \cite{17} the N400 modulation was not reported since they used different inverted faces for each icon, similarly in \cite{63} the familiarity with the face being presented as stimulus significantly produced stronger face specific ERPs. The short timing might also affect the N400 and causes conflicting components to overlap and hide it. 
Except for the P300, all latencies of ERP components were longer than reported in similar studies on colored and inverted faces. In \cite{19} where a green face acted as a stimulus the latencies were earlier than the inverted face ones \cite{17}, here, the latencies were even late. The first study employs a SOA of 250 ms whereas the second, uses a SOA of 180 ms. In the present study we fixed an SOA of 110 ms. The delay is primarily an effect of fast stimulation as demonstrated in works studying the variation of fast stimulation and ISI \cite{61, 64}, they found that the more the stimulation is faster, the more the latencies will be delayed.  
The online control results demonstrate the feasibility of single-trial control in fast SOA with strong ERPs in the time window of 100-500 ms post-stimulus, however, the users were separated in two highly variable categories in term of performance, the highly performing subjects reach near perfect control, on the contrary, weakly performing subjects do not reach the 70\% \cite{65} minimum rate required for control. Therefore the ideal SOA remains subject specific \cite{61}. 
	
Although the difference among classifiers was not significant, a trend emerges for the superiority of neural networks outperforming state-of-the art methods even in a small size data settings. This context favors compact models with a modest number of trainable parameters and medium capacity over larger and wider models. Another benefit of such size lies in maintaining both short calibration and model training times, as the signal acquisition session’s duration was around 5 minutes and the best performing model took around a minute and half for training.

A real-time control requires low processing latency, large models violate this property on regular computers with conventional compute lacking dedicated acceleration hardware like GPUs. The MAC and inference time exhibited by EEGTCNet puts it at the top among other models with variable size and capacity.

The generalization of our results are limited by a set of multiple factors, as in most BCI studies the tested user group contains only healthy subjects, although face stimuli demonstrated an effective prevention against ERP response decline in the target user group i.e. patients with neurodegenerative disease \cite{66, 67}, the performance in similar fast SOA remains unexplored as previous studies with fast SOAs reported experiments with regular flashing paradigms \cite{68}. Additionally, third of the participants could not reach the control threshold, a higher percentage than the reported 11\% (6 out of 54 subjects) in the only open dataset available using face stimulus in an ERP based BCI, further investigations with a larger population size is required to mitigate this problem.
	
The robustness of deep neural networks against day-to-day session variability is still to be assessed since the experiment was conducted in the same day. Also the type of electrodes used for EEG acquisition, the higher instability of dry electrodes in maintaining decent signal quality remains the most vulnerable piece in the non-invasive based BCIs, we were cautious in avoiding to use automatic artifact rejection (AAR) methods, as it turned out that different AAR techniques harms the classification of ERPs \cite{69}. 

Despite the harmful competitive effect to N170 between orientation and color revealed in \cite{59}, the experiment tested a single color and the effect of different colors coupled with inversion is still yet to be determined, moreover, in \cite{20} the difference between the red and green colors was not significant, furthermore, the colors red, green and magenta combined with different shapes each, had the highest detection rate in the gaze independent center-speller \cite{70}. The findings from these papers suggest the use of multiple colors with the face stimulus in the speller. We see this line of research promising for our study improvement. 

Cross subject model training uses data from different subjects to construct a large general model that can classify trials from a given subject’s unseen data in the training set successfully \cite{71}, in fact DeepConvNet had similar performance to EEGNet in all cross subject evaluations in \cite{28}, also EEGInception outperformed all models when trained with a relativity huge dataset \cite{54}. The bigger size of datasets allows the training of higher capacity models and solves the overfitting problem due to very small trainable parameters per training samples. As we had shown using large models is unlikely in similar context to our study, thus model compression and optimization for inference are required to produce an identically accurate yet faster model, this can be achieved through a plethora of techniques stemming from the deep learning literature, from knowledge distillation approach \cite{72} where two models teacher-student are trained to achieve similar performance while the student is much smaller than the teacher, to mixed precision training \cite{73} , model pruning \cite{74} and quantization aware training \cite{75}.

\section{Conclusion}
The present work, aimed at building a fast and accurate ERP based BCI to operate with more practical equipment using Dry electrodes. Through the optimization of every block in the system, first, we combined the face inversion stimulation technique with chromatic difference (Red colored stimulus) to elicit stronger ERP components in fast stimulation settings, then, to combat the inherited limitations of EEG signals, we tested state-of-the art deep learning methods which proved efficient in correctly classifying the online single trial control commands. In future work, we see two major improvements, one in the stimulus and the other in training the neural networks. The small number of commands allows for adding more colors for each row or column in order to boost the subject attention. Furthermore, the success of neural networks in decoding highly variant data opens the door for training subject independent classifiers and reduce or suppress the calibration phase altogether. We will explore these lines of work for more robust BCI systems.

\section{Acknowledgments}
The authors would like to express their deep gratitude for the authors of open access papers and for the developers of the open source code used throughout this study.

\bibliographystyle{IEEEtran}
\bibliography{erp_pilot_ref}

\section*{Appendix}

\subsection{Architectures of neural networks trained on our ERP dataset}
\captionsetup{labelformat=AppendixTables}
\setcounter{table}{0}

\begin{table}[ht!]
\centering
\begin{tabular}{|l|p{1cm}|l|p{2cm}|l|l|p{1.6cm}|} 
\hline
Layer           & Number filters & Size   & Number params & Output & Activation & options    \\ \hline
Input           &                & 15x205 &               &        &            &            \\ \hline
Zero Padding 1D &                &        &               & 15x213 &            & Padding=4  \\ \hline
SeparableConv1d & 4              & 16     & 315           & 4x25   &            & stride = 8 \\ \hline
Activation      &                &        &               & 4x25   & tanh       &            \\ \hline
Flatten         &                &        &               & 100    &            &            \\ \hline
Dense           & 1              &        & 101           & 1      & Sigmoid    &            \\ \hline
\end{tabular}
\caption{Architecture of SepConv1D \cite{52}}
\label{tab:my-table}
\end{table}

\begin{table}[h]
\centering
\begin{tabular}{|l|p{2cm}|l|p{2cm}|l|l|p{2.cm}|} 
\hline
Layer           & Number filters & Size   & Number params & Output   & Activation & options                                                                                                          \\ 
\hline
Input           & ~              & 15x205 & ~             & ~        & ~          & ~                                                                                                                \\ 
\hline
Reshape         & ~              & ~      & ~             & 1x15x205 & ~          & ~                                                                                                                \\ 
\hline
Conv2D          & 8              & 1x32   & 256           & 8x15x205 & Linear     & \begin{tabular}[c]{@{}l@{}}Padding = \\ same,  \\bias = False\end{tabular}                                          \\ 
\hline
BatchNorm       & ~              & ~      & 16            & ~        & ~          & ~                                                                                                                \\ 
\hline
DepthwiseConv2D & 16             & 15x1   & 240           & 16x1x205 & Linear     & \begin{tabular}[c]{@{}l@{}}Bias = False, \\Depth\\ multiplier\\ = 2,  \\Max weight\\ norm \\ constraint\\ = 1\end{tabular}  \\ 
\hline
BatchNorm       & ~              & ~      & 32            & 16x1x205 & ~          & ~                                                                                                                \\ 
\hline
Activation      & ~              & ~      & ~             & 16x1x205 & ELU        & ~                                                                                                                \\ 
\hline
AveragePool2D   & ~              & 1x4    & ~             & 16x1x51  & ~          & ~                                                                                                                \\ 
\hline
Dropout         & ~              & ~      & ~             & 16x1x51  & ~          & P = 0.5                                                                                                          \\ 
\hline
SeparableConv2D & 16             & 1x16   & 512           & 16x1x51  & Linear     & \begin{tabular}[c]{@{}l@{}}Padding = \\ same, \\Bias = False\end{tabular}                                           \\ 
\hline
BatchNorm       & ~              & ~      & 32            & 16x1x51  & ~          & ~                                                                                                                \\ 
\hline
Activation      & ~              & ~      & ~             & 16x1x51  & ELU        & ~                                                                                                                \\ 
\hline
AveragaePool2D  & ~              & 1x8    & ~             & 16x1x6   & ~          & ~                                                                                                                \\ 
\hline
Dropout         & ~              & ~      & ~             & 16x1x6   & ~          & P = 0.5                                                                                                          \\ 
\hline
Flatten         & ~              & ~      & ~             & 96       & ~          & ~                                                                                                                \\ 
\hline
Dense           & 1              & ~      & 97            & 1        & Sigmoid    & Max norm constraint = 0.25                                                                                       \\
\hline
\end{tabular}
\caption{Architecture of EEGNet \cite{28}}
\end{table}

\begin{table}[ht!]
\centering
\begin{tabular}{|l|p{2cm}|l|p{2cm}|l|l|p{2.cm}|} 
\hline
Layer           & Number filters & Size   & Number params & Output   & Activation & options                                                                                                          \\ 
\hline
Input           & ~              & 15x205 & ~             & ~        & ~          & ~                                                                                                                \\ 
\hline
Reshape         & ~              & ~      & ~             & 1x15x205 & ~          & ~                                                                                                                \\ 
\hline
Conv2D          & 8              & 1x32   & 256           & 8x15x205 & Linear     & Padding = same, bias = False                                                                                     \\ 
\hline
BatchNorm       & ~              & ~      & 16            & ~        & ~          & ~                                                                                                                \\ 
\hline
DepthwiseConv2D & 16             & 15x1   & 240           & 16x1x205 & Linear     & \begin{tabular}[c]{@{}l@{}}Bias = False, \\Depth\\ multiplier = \\2,  \\Max weight\\ norm \\ constraint \\ = 1\end{tabular}  \\ 
\hline
BatchNorm       & ~              & ~      & 32            & 16x1x205 & ~          & ~                                                                                                                \\ 
\hline
Activation      & ~              & ~      & ~             & 16x1x205 & ELU        & ~                                                                                                                \\ 
\hline
AveragePool2D   & ~              & 1x8    & ~             & 16x1x25  & ~          & ~                                                                                                                \\ 
\hline
Dropout         & ~              & ~      & ~             & 16x1x25  & ~          & P = 0.5                                                                                                          \\ 
\hline
SeparableConv2D & 16             & 1x16   & 512           & 16x1x25  & Linear     & Padding = same, Bias = False                                                                                     \\ 
\hline
BatchNorm       & ~              & ~      & 32            & 16x1x25  & ~          & ~                                                                                                                \\ 
\hline
Activation      & ~              & ~      & ~             & 16x1x25  & ELU        & ~                                                                                                                \\ 
\hline
AveragaePool2D  & ~              & 1x8    & ~             & 16x1x3   & ~          & ~                                                                                                                \\ 
\hline
Dropout         & ~              & ~      & ~             & 16x1x3   & ~          & P = 0.5                                                                                                          \\ 
\hline
TCN             & 12-12-12       & 4-4-1  & 1401          & 12x3     & ~          & \begin{tabular}[c]{@{}l@{}}Padding =\\ causal, \\Dilation \\rate = 1, \\Padding = 3,\\ P = 0.2\end{tabular}            \\ 
\hline
TCN             & 12-12          & 4-4    & 1224          & 12       & ~          & \begin{tabular}[c]{@{}l@{}}Padding =\\ causal, \\Dilation\\ rate = 2, \\Padding = 6,\\ P = 0.2\end{tabular}            \\ 
\hline
Dense           & ~              & ~      & 13            & 1        & Sigmoid    & ~                                                                                                                \\
\hline
\end{tabular}
\caption{Architecture of EEGTCNet  \cite{53}}
\end{table}

\begin{table}[h]
\centering
\begin{tabular}{|l|l|l|l|l|} 
\hline
Layer      & Number filters & Size & Activation & options                                          \\ 
\hline
Conv1D     & 12             & 4    & Linear     & Padding = causal, Dilation rate= 1, Padding = 3  \\ 
\hline
Chomp1D    & ~              & ~    & ~          & Padding = 3                                      \\ 
\hline
BatchNorm  & ~              & ~    & ~          & ~                                                \\ 
\hline
Activation & ~              & ~    & ELU        & ~                                                \\ 
\hline
Dropout    & ~              & ~    & ~          & P = 0.2                                          \\ 
\hline
Conv1D     & 12             & 4    & Linear     & Padding = causal, Dilation rate= 1, Padding = 3  \\ 
\hline
Chomp1D    & ~              & ~    & ~          & Padding=3                                        \\ 
\hline
BatchNorm  & ~              & ~    & ~          & ~                                                \\ 
\hline
Activation & ~              & ~    & ELU        & ~                                                \\ 
\hline
Dropout    & ~              & ~    & ~          & P = 0.2                                          \\ 
\hline
Conv1D     & 12             & 1    & Linear     & (Optional)                                       \\ 
\hline
Add        & ~              & ~    & ~          & ~                                                \\ 
\hline
Activation & ~              & ~    & ELU        & ~                                                \\
\hline
\end{tabular}
\caption{TCN block \cite{53}}
\end{table}

\begin{table}[h]
\centering
\begin{tabular}{|p{2cm}|l|p{1.1cm}|l|p{1cm}|l|l|l|p{1.5cm}|} 
\hline
Block            & name & Number filters & Size   & Number params & Output   & Activation & options    & Connected to  \\ 
\hline
Input            & ~    & ~              & 15x205 & ~             & ~        & ~          & ~          & C1, C2, C3    \\ 
\hline
Conv2D           & C1   & 8              & 1x64   & 536           & 8x15x205 & ELU        & N=4        & D1            \\ 
\hline
Depthwise Conv2D & D1   & 2              & 8x1    & 272           & 16x1x205 & ELU        & -          & N1            \\ 
\hline
Conv2D           & C2   & 8              & 1x32   & 280           & 8x15x205 & ELU        & N=2        & D2            \\ 
\hline
Depthwise Conv2D & D2   & 2              & 8x1    & 272           & 16x1x205 & ELU        & -          & N1            \\ 
\hline
Conv2D           & C3   & 8              & 1x16   & 152           & 8x15x205 & ELU        & N=1        & D3            \\ 
\hline
Depthwise Conv2D & D3   & 2              & 8x1    & 272           & 16x1x205 & ELU        & -          & N1            \\ 
\hline
Concatenate      & N1   & ~              & ~      & ~             & 48x1x205 & ~          & ~          & A1            \\ 
\hline
AveragePool 2D   & A1   & ~              & 1x4    & ~             & 48x1x51  & ~          & Stride=1x4 & C4, C5, C6    \\ 
\hline
Conv2D           & C4   & 8              & 1x16   & 6168          & 8x1x51   & ELU        & N=1        & N2            \\ 
\hline
Conv2D           & C5   & 8              & 1x8    & 3096          & 8x1x51   & ELU        & N = 0.5    & N2            \\ 
\hline
Conv2D           & C6   & 8              & 1x4    & 1560          & 8x1x51   & ELU        & N = 0.25   & N2            \\ 
\hline
Concatenate      & N2   & ~              & ~      & ~             & 24x1x51  & ~          & ~          & A2            \\ 
\hline
AveragePool 2D   & A2   & ~              & 1x2    & ~             & 24x1x25  & ~          & Stride=1x2 & C7            \\ 
\hline
Conv2D           & C7   & 12             & 1x8    & 2328          & 12x1x25  & ELU        & N = 0.5    & A3            \\ 
\hline
AveragePool 2D   & A3   & ~              & 1x2    & ~             & 12x1x12  & ~          & Stride=1x2 & C8            \\ 
\hline
Conv2D           & C8   & 6              & 1x4    & 300           & 6x1x12   & ELU        & N = 0.25   & A4            \\ 
\hline
AveragePool 2D   & A4   & ~              & 1x2    & ~             & 6x1x6    & ~          & Stride=1x2 & Dense         \\ 
\hline
Dense            & ~    & 1              & ~      & 37            & ~        & Sigmoid    & ~          & ~             \\
\hline
\end{tabular}
\caption{Architecture of EEG-Inception \cite{54}}
\end{table}

\begin{table}[h]
\centering
\begin{tabular}{|l|l|l|l|l|} 
\hline
Layer      & Number filters & Size     & Activation & options        \\ 
\hline
Conv2D     & 8              & 1x(16*N) & Linear     & Padding= same  \\ 
\hline
BatchNorm  & ~              & ~        & ~          & ~              \\ 
\hline
Activation & ~              & ~        & ELU        & ~              \\ 
\hline
Dropout    & ~              & ~        & ~          & P = 0.2        \\
\hline
\end{tabular}
\caption{EEG-Inception Convolution Block \cite{54}}
\end{table}

\begin{table}[h]
\centering
\begin{tabular}{|l|l|l|l|l|} 
\hline
Layer            & Depth & Size & Activation & options        \\ 
\hline
Depthwise Conv2D & 2     & 8x1  & Linear     & Padding=valid  \\ 
\hline
BatchNorm        & ~     & ~    & ~          & ~              \\ 
\hline
Activation       & ~     & ~    & ELU        & ~              \\ 
\hline
Dropout          & ~     & ~    & ~          & P = 0.2        \\
\hline
\end{tabular}
\caption{EEG-Inception Depthwise Convolution Block \cite{54}}
\end{table}

\clearpage
\begin{table}[ht!]
\centering
\begin{tabular}{|l|p{2cm}|l|p{2cm}|l|l|p{2cm}|} 
\hline
Layer      & Number filters & Size                                                               & Number params & Output    & Activation & options                          \\ 
\hline
Input      & ~              & 15x205                                                             & ~             & ~         & ~          & ~                                \\ 
\hline
Reshape    & ~              & ~                                                                  & ~             & 1x15x201  & ~          & ~                                \\ 
\hline
Conv2D     & 25             & 1x5                                                                & 125           & 25x15x201 & Linear     & Max norm = 2                     \\ 
\hline
Conv2D     & 25             & 15x1                                                               & 9375          & 25x1x201  & Linear     & Max norm = 2                     \\ 
\hline
BatchNorm  & ~              & ~                                                                  & 50            & 25x1x201  & ~          & Epsilon = 1e-5 , momentum = 0.1  \\ 
\hline
Activation & ~              & ~                                                                  & ~             & 25x1x201  & ELU        & ~                                \\ 
\hline
MaxPool2D  & ~              & \begin{tabular}[c]{@{}l@{}}1x2 pool size \\1x2 stride\end{tabular} & ~             & 25x1x100  & ~          & ~                                \\ 
\hline
Dropout    & ~              & ~                                                                  & ~             & 25x1x100  & ~          & P = 0.5                          \\ 
\hline
Conv2D     & 50             & 1x5                                                                & 6250          & 50x1x96   & Linear     & Max norm = 2                     \\ 
\hline
BatchNorm  & ~              & ~                                                                  & 100           & 50x1x96   & ~          & Epsilon = 1e-5 , momentum = 0.1  \\ 
\hline
Activation & ~              & ~                                                                  & ~             & 50x1x96   & ELU        & ~                                \\ 
\hline
MaxPool2D  & ~              & \begin{tabular}[c]{@{}l@{}}1x2 pool size \\1x2 stride\end{tabular} & ~             & 50x1x48   & ~          & ~                                \\ 
\hline
Dropout    & ~              & ~                                                                  & ~             & 50x1x48   & ~          & P = 0.5                          \\ 
\hline
Conv2D     & 100            & 1x5                                                                & 25000         & 100x1x44  & Linear     & Max norm = 2                     \\ 
\hline
BatchNorm  & ~              & ~                                                                  & 200           & 100x1x44  & ~          & Epsilon = 1e-5 , momentum = 0.1  \\ 
\hline
Activation & ~              & ~                                                                  & ~             & 100x1x44  & ELU        & ~                                \\ 
\hline
MaxPool2D  & ~              & \begin{tabular}[c]{@{}l@{}}1x2 pool size \\1x2 stride\end{tabular} & ~             & 100x1x22  & ~          & ~                                \\ 
\hline
Dropout    & ~              & ~                                                                  & ~             & 100x1x22  & ~          & P = 0.5                          \\ 
\hline
Conv2D     & 200            & 1x5                                                                & 100000        & 200x1x18  & Linear     & Max norm = 2                     \\ 
\hline
BatchNorm  & ~              & ~                                                                  & 400           & 200x1x18  & ~          & Epsilon = 1e-5 , momentum = 0.1  \\ 
\hline
Activation & ~              & ~                                                                  & ~             & 200x1x18  & ELU        & ~                                \\ 
\hline
MaxPool2D  & ~              & \begin{tabular}[c]{@{}l@{}}1x2 pool size \\1x2 stride\end{tabular} & ~             & 200x1x9   & ~          & ~                                \\ 
\hline
Dropout    & ~              & ~                                                                  & ~             & 200x1x9   & ~          & P = 0.5                          \\ 
\hline
Flatten    & ~              & ~                                                                  & ~             & 1800      & ~          & ~                                \\ 
\hline
Dense      & 1              & ~                                                                  & 1801          & 1         & Sigmoid    & Max norm = 0.5                   \\
\hline
\end{tabular}
\caption{Architecture of DeepConvNet \cite{27}}
\end{table}

\end{document}